\documentclass[11pt]{article}
\usepackage{moriond,epsfig,subfigure,axodraw}
\usepackage{amsmath,amssymb}
\usepackage{epsfig,graphicx}

\def\be{\begin{equation}}
\def\ee{\end{equation}}
\def\bea{\begin{eqnarray}}
\def\eea{\end{eqnarray}}

\newcommand{\fssd}[1]{#1\!\!\!\!/}

\begin{document}
\hspace{12cm} IPPP $\,\,\,$07/24

\hspace{12cm} DCPT 07/48 \\[-0.9cm]
\vspace*{3cm}
\title{Photons as a Probe of Minicharged Particles}
\author{Joerg Jaeckel }

\address{Institute for Particle Physics Phenomenology and Centre for Particle Theory, \\
University of Durham, Durham, DH1 3LE, UK}

\maketitle\abstracts{ Low energy experiments with photons can
provide deep insights into fundamental physics. In this note we
concentrate on minicharged particles. We discuss how they can arise
in extensions of the standard model and how we can search for them
using a variety of laboratory experiments. }

\section{Introduction -- Light particles coupled to photons}
Light particles weakly coupled to photons appear in a variety of
extensions of the standard model. A prominent example is the axion
invented to solve the strong CP problem
\cite{Peccei:1977hh,Weinberg:1977ma,Wilczek:1977pj}. The axion is an
example of a (pseudo-)scalar particle $\phi$ coupled to two photons
via a dimension five interaction,
\begin{equation}
\label{scalarinteraction}
{\mathcal L}^{(-)}_{\rm{int}}
  =-\frac{1}{4}g\phi^{(-)}F_{\mu\nu}\widetilde{F}^{\mu\nu}
  =g\phi^{(-)}(\vec{E}\cdot\vec{B}),
\end{equation}
where the coupling constant $g$ has dimensions $1/{\rm{Mass}}$. Other examples of such light spin-$0$ bosons are
familons~\cite{Wilczek:1982rv}, Majorons~\cite{Chikashige:1980ui,Gelmini:1980re}, the dilaton, and moduli,
to name just a few\footnote{For a scalar particle the
$\tilde{F}$ in Eq. \eqref{scalarinteraction} has to be replaced by an $F$.}. Independent of their origin
light particles coupled to two photons as in Eq. \eqref{scalarinteraction} are often called axion-like particles or ALPs.

ALPs can be constrained by a variety of astrophysical
observations \cite{Frieman:1987ui,Raffelt:1985nk,Raffelt:1987yu,Raffelt:1996}.
However, these bounds can be avoided in more complicated models \cite{Masso:2005ym,Jain:2005nh,Jain:2006ki,Masso:2006gc,Mohapatra:2006pv,Jaeckel:2006id,Brax:2007ak}
making it desirable to have clean and controlled laboratory tests \cite{Jaeckel:2006id}.

\begin{figure}
\vspace{-1.3cm}
\subfigure{ \scalebox{1}[1]{
\begin{picture}(200,120)(0,0)
\includegraphics[bbllx=99,bblly=35,bburx=1173,bbury=410,width=.4\textwidth]{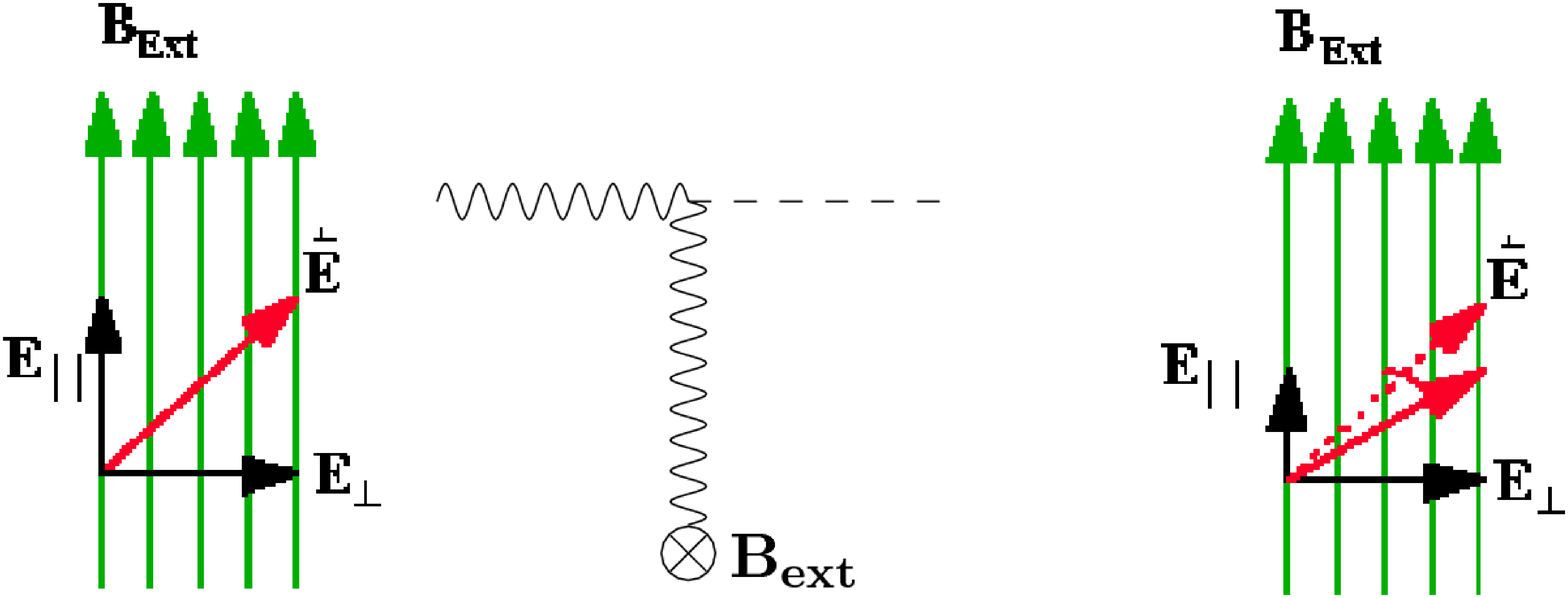}
\Text(-167,-10)[c]{\scalebox{1.}[1.]{$\mathbf{before}$}}
\Text(-13,-10)[c]{\scalebox{1.}[1.]{$\mathbf{after}$}}
\end{picture}}
\label{alpconversion}}
\hspace{1cm}
\subfigure{
\scalebox{1}[1]{
\begin{picture}(200,120)(0,0)
\includegraphics[bbllx=100,bblly=85,bburx=1171,bbury=465,width=.4\textwidth]{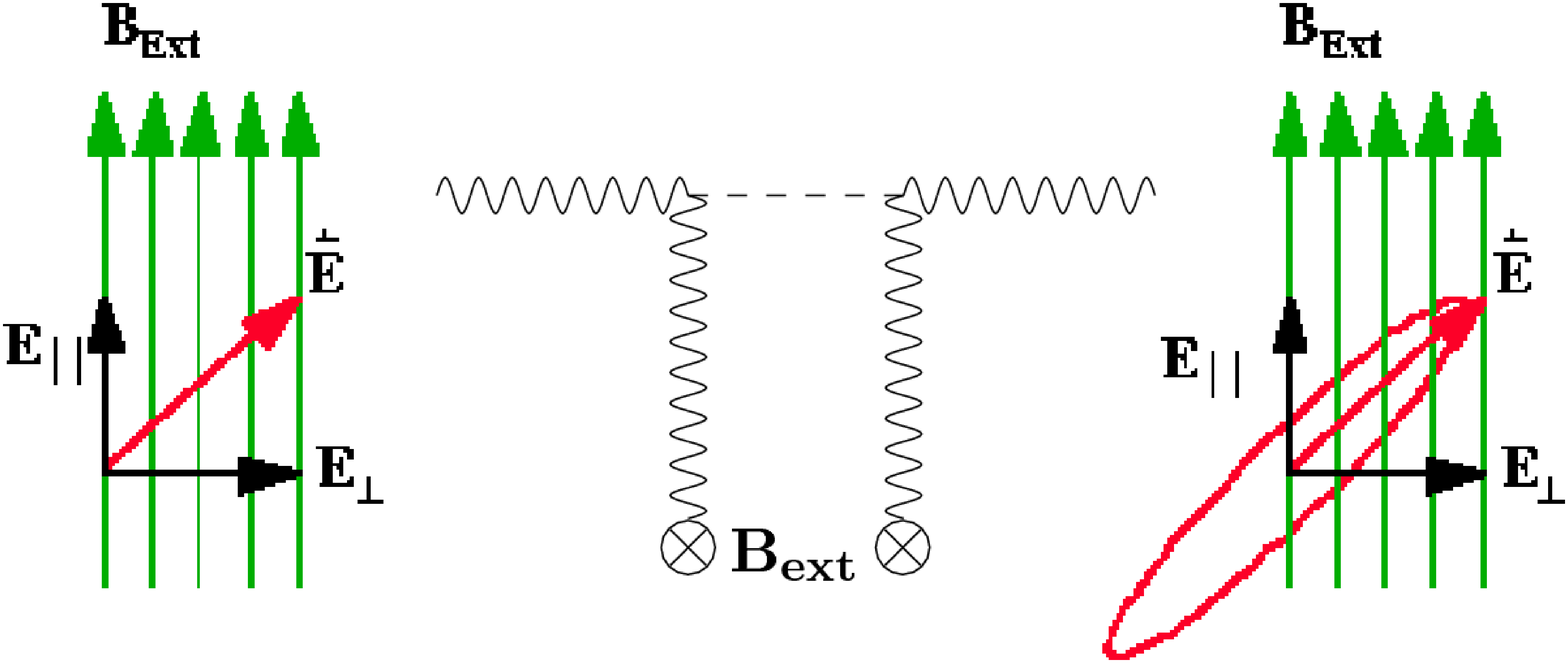}
\Text(-167,-10)[c]{\scalebox{1.}[1.]{$\mathbf{before}$}}
\Text(-13,-10)[c]{\scalebox{1.}[1.]{$\mathbf{after}$}}
\end{picture}}
\label{alpconversion}}
\vspace{0.3cm}
\caption{Rotation (left) and ellipticity (right) caused by the existence of a light neutral spin-0 boson (adapted from$^{\ref{brandi}}$).
In a homogeneous magnetic background $\vec{B}$, the interaction Eq. \eqref{scalarinteraction} can convert the laser photons into pseudoscalars
(left).
The leading order contribution to this process comes from the
term $\sim \vec{E}_{\gamma}\cdot\vec{B}$.
The polarization of a photon is given by the direction of the
electric field of the photon, $\vec{E}_\gamma$.
Therefore, only those fields polarized
parallel to the background magnetic field will have nonvanishing
$\vec{E}_{\gamma}\cdot\vec{B}\neq0$ and interact with the
\mbox{pseudoscalar} particles.
Virtual particle production (right) leads to phase shift and in turn an ellipticity. Very naively speaking the virtual intermediate particle
is massive and therefore a bit slower than the massless photon.
}
\label{alp}
\end{figure}

Two types of experiments are particularly noteworthy. First there are experiments that look for changes in the polarization when a (linear) polarized laser
beam passes through a strong magnetic field as depicted in Fig. \ref{alp}. This is a disappearance experiment where the produced particles are not detected.
A pioneering experiment of this type was done by the BFRT collaboration \cite{Semertzidis:1990qc,Cameron:1993mr} and produced limits on the allowed couplings and masses.
Recently,
the PVLAS experiment reported the observation of a non-vanishing rotation signal \cite{Zavattini:2005tm}. This has sparked a significant amount of theoretical
\cite{Masso:2005ym,Jain:2005nh,Jain:2006ki,Masso:2006gc,Mohapatra:2006pv,Jaeckel:2006id,Brax:2007ak,Gies:2006ca,Gies:2006hv,Abel:2006qt,Ahlers:2006iz,Jaeckel:2007pv,Foot:2007cq,Melchiorri:2007sq,Kim:2007wj} work
as well as planning and construction of new
experiments \cite{Rabadan:2005dm,Pugnat:2005nk,Rizzo:Patras,Ehret:2007cm,Baker:Patras,Cantatore:Patras,Ringwald:2006rf,Battesti:2007um}. The Q\&A experiment has already
published some data \cite{Chen:2006cd} but its sensitivity is not yet sufficient to test PVLAS.

\begin{figure}
\begin{center}
\includegraphics[width=8.5cm]{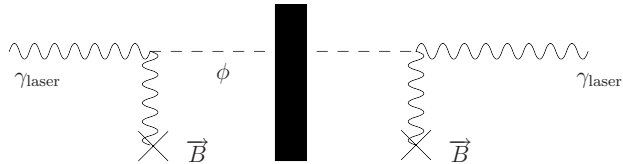}
\end{center}
\vspace{-0.5cm}
\caption{Schematic view of a ``light shining through a wall'' experiment.
(Pseudo-)scalar production through photon conversion in a magnetic field (left), subsequent travel
through a wall, and final detection through photon regeneration
(right). }
\label{fig:ph_reg}
\end{figure}

Second, there are photon regeneration experiments or ``light shining through walls'' experiments, as shown in Fig. \ref{fig:ph_reg}, where the produced particles are
reconverted into photons and then detected \cite{Anselm:1986gz,Gasperini:1987da,VanBibber:1987rq,Ringwald:2003ns,Sikivie:2007qm}. BFRT also run a setup of
this type \cite{Ruoso:1992nx,Cameron:1993mr}. And, particularly interesting, most upcoming
experiments will be of this type (or at least have a ``light shining through walls'' stage)~\cite{Pugnat:2005nk,Rizzo:Patras,Ehret:2007cm,Baker:Patras,Cantatore:Patras}.

ALPs have zero electric charge. What about light \emph{charged} particles? At first one might be tempted to exclude this possibility simply by saying
if it is charged and lighter than an electron it is excluded by a huge number of experiments. However, this is implicitly based on the assumption that charge is quantized
and the smallest quantum is not much smaller than the charge of the electron. Although strong bounds on the charges of neutrons, atoms and molecules suggest the
idea that charge quantization is a fundamental principle one needs physics beyond the standard model to enforce charge quantization \cite{Foot:1990mn}. One possibility
would be the existence of magnetic monopoles as demonstrated by Dirac's seminal argument \cite{Dirac:1931kp}. However, many extensions of the standard model do indeed
contain particles with small electric charges \cite{Ignatiev:1978xj,Okun:1983vw,Holdom:1985ag,Dienes:1996zr,Abel:2003ue,Batell:2005wa,Abel:2006qt}.

The interaction for such particles is the standard minimal coupling but with a small fraction $\epsilon$ of a unit electric charge. For example for Dirac fermions it reads,
\begin{equation}
\label{chargeinteraction}
{\mathcal L}_{\rm int}^{\rm Dsp}= \epsilon\, e\,
\overline{\psi}_{\epsilon}
\gamma_\mu \psi_{\epsilon} A^\mu.
\end{equation}

As we will see in Sect. \ref{experiments} such an interaction can be tested in optical experiments (cf. Fig.~\ref{milli}) as well as in experiments with strong
electric fields as shown in Fig.~\ref{acdc}.

\section{Minicharged particles in paraphoton models}\label{millicharge}
Minicharged particles arise most naturally in models with extra U(1) gauge degrees of freedom~\cite{Holdom:1985ag} so called paraphotons.
In this section we briefly review how kinetic mixing leads to minicharged particles.

Let us look at the simplest model with two U(1) gauge groups, one being our electromagnetic U(1), the other a hidden sector U(1) under which
all Standard Model particles have zero charge. The most general Lagrangian allowed by the symmetries is,
\begin{equation}
\label{kineticmixing}
{\mathcal{L}}=-\frac{1}{4} F^{\mu\nu}F_{\mu\nu}-\frac{1}{4}B^{\mu\nu}B_{\mu\nu}-\frac{1}{2}\chi\,F^{\mu\nu}B_{\mu\nu},
\end{equation}
where $F^{\mu\nu}$ is the field strength tensor for the ordinary electromagnetic U(1) gauge field $A^{\mu}$ and $B^{\mu\nu}$ is the field strength for the hidden
sector U(1) field $B^{\mu}$, i.e. the paraphoton.
The first two terms are the standard kinetic terms for the photon and paraphoton fields, respectively. Because the field strength itself is gauge invariant
for U(1) gauge fields the third term is also allowed by the gauge symmetries (and Lorentz symmetry).
This term corresponds to a non-diagonal kinetic term, i.e. a so called kinetic mixing.


The kinetic term can be diagonalized by a shift
\begin{equation}
\label{shift}
B^{\mu}\rightarrow \tilde{B}^{\mu}-\chi A^{\mu}.
\end{equation}
Aside from a multiplicative renormalization of the gauge coupling $e^2\rightarrow e^2/(1+\chi^2)$ the visible sector fields remain unaffected by this shift.

Let us now assume that we have a hidden sector fermion $f$ that has charge one under $B^{\mu}$. Applying the shift Eq. \eqref{shift} to the coupling term we find,
\begin{equation}
e_{\rm{h}}\bar{f}\fssd{B} f\rightarrow e_{\rm{h}}\bar{f}\fssd{\tilde{B}} f-\chi e_{\rm{h}}\bar{f}\fssd{A} f,
\end{equation}
where $e_{\rm{h}}$ is the hidden sector gauge coupling.
We can read off that the hidden sector particle now has a charge
\begin{equation}
\label{epsiloncharge}
\epsilon e=-\chi e_{\rm{h}}
\end{equation}
under the visible electromagnetic gauge field $A^{\mu}$ which has gauge coupling $e$. Since $\chi$ is an arbitrary number  the fractional electric charge $\epsilon$
of the hidden sector fermion $f$ is not necessarily integer.

For small $\chi\ll1$
\begin{equation}
|\epsilon|\ll 1
\end{equation}
and $f$ becomes a minicharged particle. From now on we will concentrate on this case\footnote{Light particles with charge $\epsilon={\mathcal{O}}(1)$
are excluded by experiments and very massive particles give negligible contributions in experiments such as PVLAS or the upcoming optical experiments.}.

\begin{figure}
\vspace{-0.5cm}
\begin{center}
\subfigure[]{
\scalebox{1.2}[1.2]{
\includegraphics[bbllx=230,bblly=616,bburx=381,bbury=717,width=3.5cm]{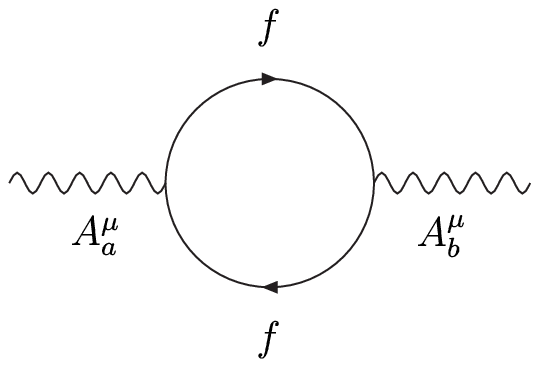}}
\label{fieldmixing}}
\hspace{1cm}
\subfigure[]{
\scalebox{1.}[1.]{
\includegraphics[bbllx=243,bblly=616,bburx=380,bbury=708,width=3.6cm]{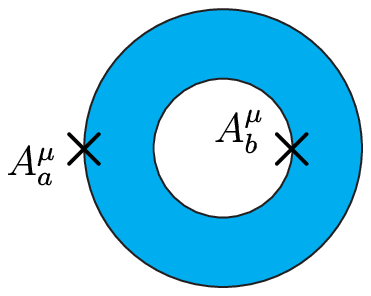}
}\label{stringmixing}}
\hspace{1cm}
\subfigure[]{
\scalebox{1.}[1.]{
\includegraphics[bbllx=114,bblly=105,bburx=527,bbury=574,width=3cm]{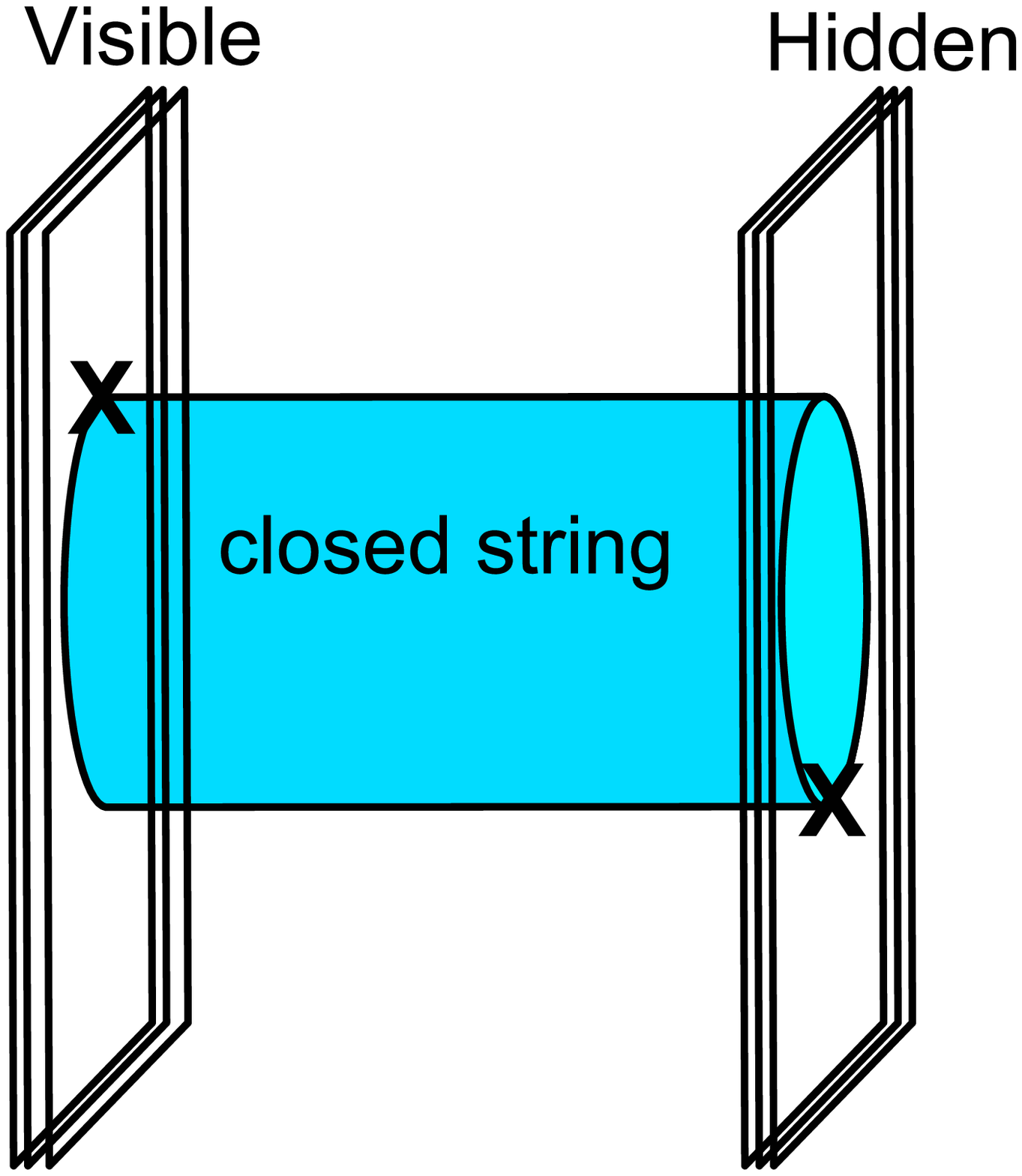}
}\label{stringmixing2}}
\end{center}
\vspace*{-0.4cm}
\caption[...]{(a) One-loop diagram which contributes to kinetic-mixing in field theory,
 (b) its equivalent in open string theory and (c) reinterpretation of (b) as a closed string exchange in the context of D-brane models.
\label{loop}}
\end{figure}

To conclude this section let us comment on the origin of the kinetic mixing term in Eq.~\eqref{kineticmixing}
(for more details see, e.g., \cite{Holdom:1985ag,Dienes:1996zr,Abel:2003ue,Abel:2006qt}). First of all it should be stressed
that the kinetic mixing term is allowed by all symmetries and therefore a free parameter from the viewpoint of an effective low energy field theory.
Having said this it is also clear that such a term will typically be generated by loop diagrams in quantum field theory. For example if we have a heavy particle
that is charged under both the electromagnetic as well as the hidden sector U(1) gauge group we find a diagram as in Fig. \ref{fieldmixing} which automatically
generates a kinetic mixing term. In string theory a similar diagram with an open string going around the loop exists (Fig.~\ref{stringmixing}).
In D-brane models of string theory stacks of D-branes generate U(N) gauge groups. The diagram Fig. \ref{stringmixing} can then be
understood as a closed string exchange between two stacks of D-branes. One may imagine that we live on a stack of such D-branes, the ``visible'' sector, that
communicates via such a closed string with another stack of D-branes, the ``hidden'' sector (cf. Fig.~\ref{stringmixing2}).
In this way observing kinetic mixing is a first step towards observing the hidden sector which is a common feature in string theory models.

\section{Searching minicharged particles in the laboratory}\label{experiments}
Let us now look how we can actually search for these minicharged particles in the laboratory\footnote{As for ALPs the astrophysical bounds
are quite strong, $\epsilon\lesssim 10^{-14}$ (see, e.g., \cite{Davidson:1991si,Mohapatra:1990vq,Mohapatra:1991as,Davidson:1993sj,Davidson:2000hf}),
but may be circumvented in some models\cite{Masso:2006gc,Foot:2007cq,Kim:2007wj} (note, however, \cite{Melchiorri:2007sq} for a bound that may be more difficult to avoid).}.

A classic probe of minicharged particles in the laboratory is the invisible decay of orthopositronium~\cite{Dobroliubov:1989mr,Mitsui:1993ha,Badertscher:2006fm}.
The current limit from this type of experiments is $\epsilon<3.4\times 10^{-5}$.

\begin{figure}
\subfigure{
\scalebox{1}[1]{
\begin{picture}(200,120)(0,0)
\includegraphics[bbllx=99,bblly=39,bburx=1166,bbury=415,width=.4\textwidth]{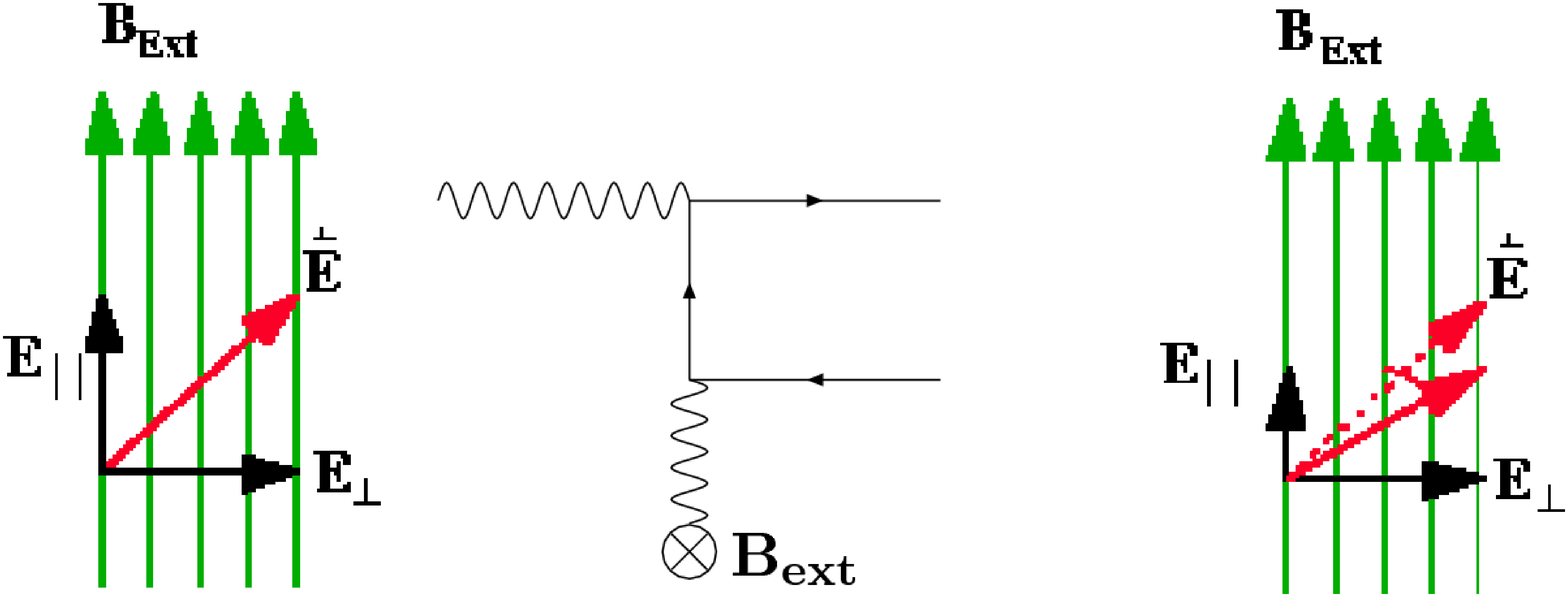}
\Text(-167,-10)[c]{\scalebox{1.}[1.]{$\mathbf{before}$}}
\Text(-13,-10)[c]{\scalebox{1.}[1.]{$\mathbf{after}$}}
\end{picture}}
\label{millirot}}
\hspace{1cm}
\subfigure{
\scalebox{1}[1]{
\begin{picture}(200,120)(0,0)
\includegraphics[bbllx=109,bblly=97,bburx=1171,bbury=478,width=.4\textwidth]{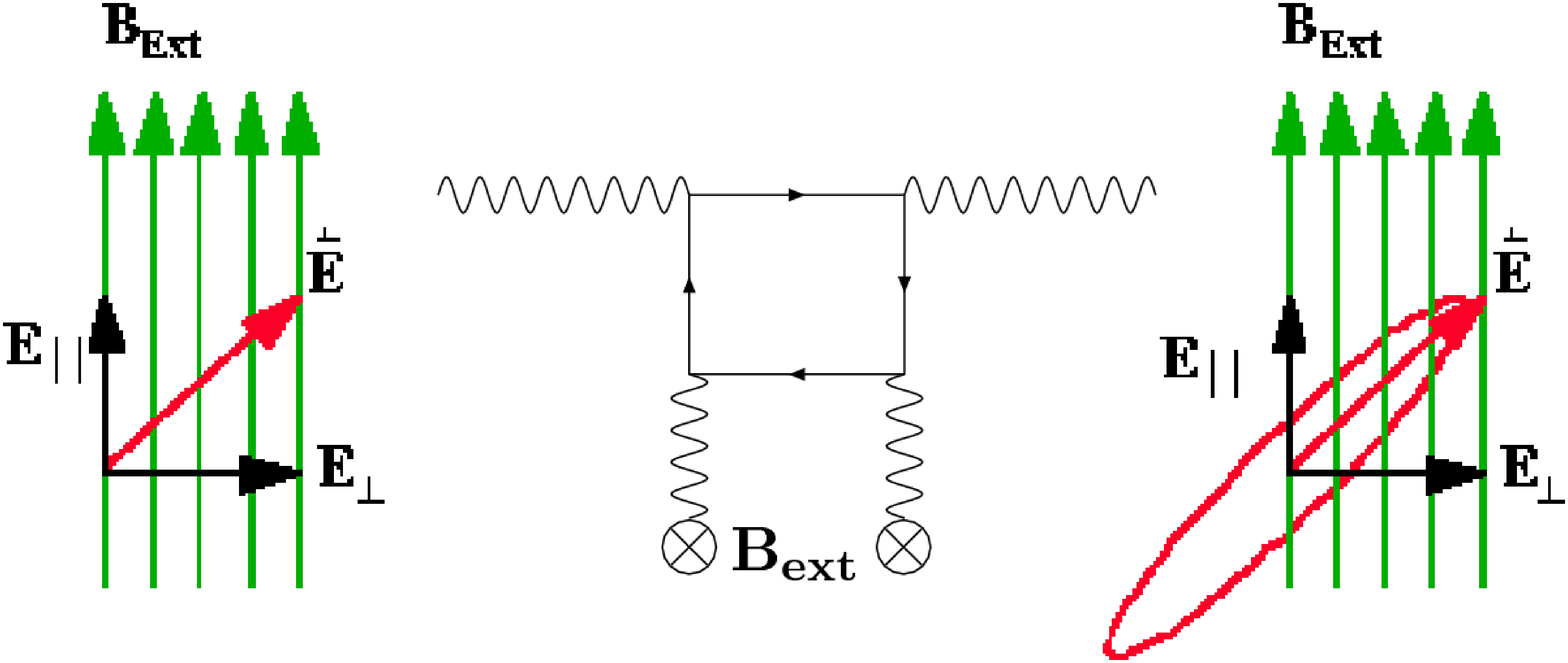}
\Text(-167,-10)[c]{\scalebox{1.}[1.]{$\mathbf{before}$}}
\Text(-13,-10)[c]{\scalebox{1.}[1.]{$\mathbf{after}$}}
\end{picture}}
\label{milliell}}
\vspace{0cm}
\caption{Rotation (left) and ellipticity (right) caused by the existence of a light minicharged particle.
In a homogeneous magnetic background $\vec{B}$, the interaction Eq. \eqref{chargeinteraction} can convert the laser photons into pairs of charged particles
(left). The conversion probability depends on the relative orientation of the magnetic field and the laser polarization, resulting in an overall
rotation. In a similar manner virtual pair production (right) in the magnetic field leads to an orientation dependent index of refraction. This causes a phase shift
that appears as an ellipticity in the outgoing beam.
}
\label{milli}
\end{figure}

In the small mass range optical experiments provide an even more powerful tool in the search for minicharged particles \cite{Gies:2006ca}.
Again we can test for changes in the
polarization of a laser beam after it has passed through a strong magnetic field. As for ALPs we can have real and virtual production of particles but now
it is pair production instead of single particle production. The relevant processes are depicted in Fig. \ref{milli}.

For small masses data from the BFRT and Q\&A experiments \cite{Cameron:1993mr,Chen:2006cd} constrain \cite{Ahlers:2006iz}
\begin{equation}
\epsilon< 1.2 \times10^{-6},\quad\,\,{\rm{for}}\quad m_{\epsilon}\lesssim 10^{-2}\,{\rm{eV}}.
\end{equation}

Interpreted as a minicharged particle effect the observed rotation in the PVLAS experiment~\cite{Zavattini:2005tm} would suggest particles
with a mass $m_{\epsilon}\lesssim 0.1\,{\rm{eV}}$ and a charge in the range $(0.7-1.2)\times10^{-6}$. This makes this interpretation testable in the immediate
future (see also below).

Another way to search for minicharged particles is to employ Schwinger pair production in strong electric fields. In a strong electric field charged
particles gain energy when they are separated along the lines of the electric field. Now, if a virtual particle-antiparticle pair generated by a vacuum fluctuation
is separated by a large enough distance such that the energy gain in the electric field is bigger than the rest mass of the particles the virtual pair becomes real.
In other words a strong electric
field can ``decay'' into particle-antiparticle pairs of charged particles. This process is similar to tunneling. An energy barrier (rest mass) with finite extent
(distance between the particles that is sufficient such that the energy gain in the electric field can compensate for the rest mass) can be quantum mechanically crossed.
As expected the rate is exponentially small if the barrier is high (large mass) and the distance is large (high mass, weak electric field and small charge).
However, if the mass is small pair production can be quite effective.

One possibility to generate such strong electric fields is to use accelerator cavities.

The remaining question is how do we know that we have produced minicharged particles in the cavity. Direct detection seems difficult because the cross section
decreases with $\sim \epsilon^2$ and becomes quite small for small $\epsilon$. Here, it is useful that once Schwinger pair production sets in it really can produce
a lot of particles. Such a massive production of particles drains a macroscopic amount of energy from the cavity which can be detected (e.g. it would lead to
a decrease in the quality factor of the cavity). From available data on the energy loss of cavities for the TESLA accelerator one can infer\cite{Gies:2006hv}
strong limits on the existence of light minicharged particles $\epsilon \lesssim 10^{-6}$ for masses $m_{\epsilon} \lesssim 10^{-3.5}\,{\rm{eV}}$.

\begin{figure}
\subfigure{
\scalebox{1}[1]{
\begin{picture}(200,100)(0,-30)
\includegraphics[bbllx=341,bblly=67,bburx=126,bbury=747,angle=-90,width=.4\textwidth]{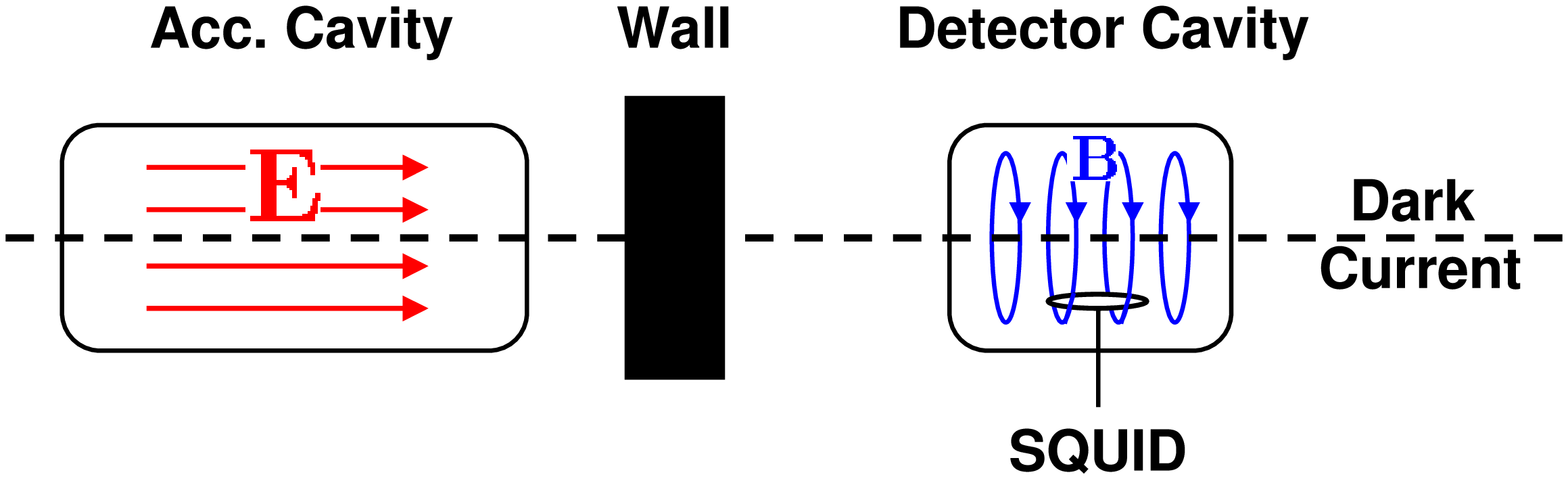}
\end{picture}}
\label{detector}}
\hspace{1cm}
\subfigure{
\scalebox{1}[1]{
\begin{picture}(200,100)(0,0)
\includegraphics[bbllx=84,bblly=4,bburx=375,bbury=182,width=.4\textwidth]{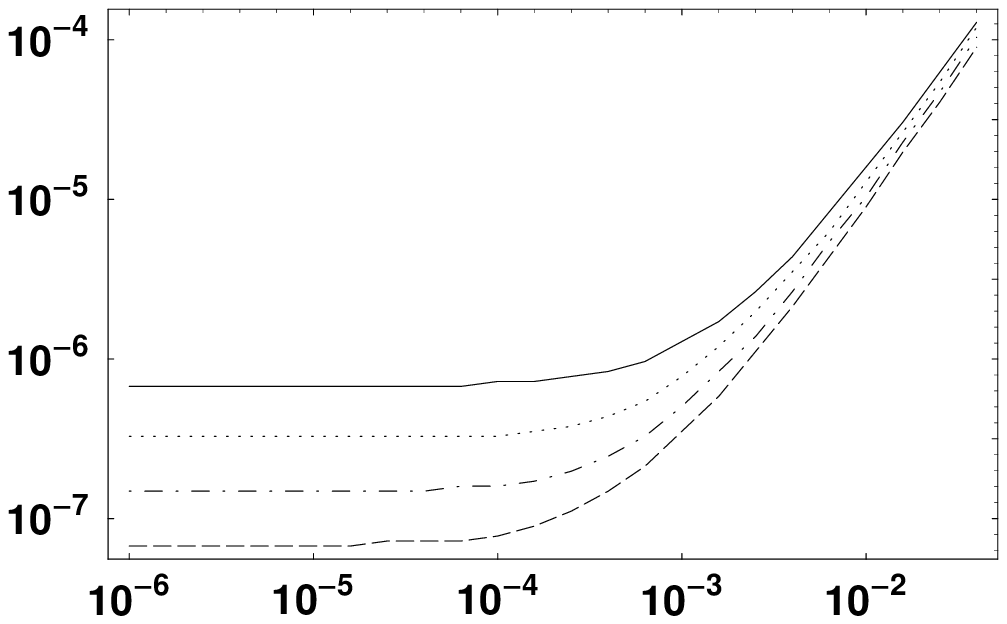}
\Text(-187,95)[c]{\scalebox{1.4}[1.4]{$\mathbf{\epsilon}$}}
\Text(-13,-10)[c]{\scalebox{1.1}[1.1]{$\mathbf{m_{\epsilon}}$}}
\end{picture}}
\label{current}}
\caption{``Dark current flowing through the walls'' experiment (left). In a cavity with strong electric fields Schwinger pair production produces pairs of minicharged
particles.
The produced particles have typically momenta along the lines of the electric field. Positive particles flying in one direction and negative particles in the opposite
direction. This results in a current. This current is, however, made up of very weakly interacting particles and can therefore pass through thick layers of
material (in contrast to, e.g., a current made up from electrons). This current can then be measured on the
other side of the wall (e.g. by measuring the generated magnetic field). For sufficiently strong electric field pair production is very efficient
and the currents can reach values measurable with current technology. The plot in the right panel shows a (very optimistic) estimate for a
current generated in a cavity of length 20 cm and radius 10 cm with a field strength of $\sim$15 MV/m $^{\ref{giesprep}}$. From top to bottom the lines
correspond to currents from $\mu A$ to $nA$.}
\label{acdc}
\end{figure}

Using Schwinger pair production process one could even set up a detection experiment (measuring the energy loss is a disappearance experiment) by
measuring currents
generated in the cavities as depicted in Fig. \ref{acdc}. With available technology such an experiment has chances to probe the interesting region of
$\epsilon \sim {\rm{few}}\times 10^{-7}$ favored by a minicharged particle interpretation of the PVLAS data.

\section{Conclusions and outlook}
Light particles coupled to photons appear in a wide variety of possible extensions of the standard model.
In particular minicharged particles can arise from models with extra U(1) gauge degrees of freedom.
Such particles can be searched for in experiments with photons as, e.g., in experiments that shine laser light through strong magnetic fields or by
searching for Schwinger pair production in strong electric fields.
Another promising approach is to search for the effects of the additional U(1) degrees of freedom\cite{ahlersprep}.

Inspired by the PVLAS observation several new experiments suitable for the search for light particles coupled to photons are in planning or are already under construction.
Additional experiments such as the ``Dark current flowing through a wall'' could be build with present technology.
These experiments will not only allow to test the PVLAS result but will probe whole classes of viable extensions of the standard model.

\section*{Acknowledgments}
The author would like to thank the organizers of the ``Rencontres de Moriond: Electroweak Interactions and Unified Theories'' for a pleasant and productive
meeting in a wonderful environment.
Furthermore, he would like to thank S.A. Abel, M. Ahlers, H.~Gies, V.V.~Khoze, E.~Masso, J.~Redondo and A. Ringwald for fruitful collaboration and many interesting discussions.

\end{document}